**Title:** Tree of life reveals clock-like speciation and diversification

S. Blair Hedges[a–c,1], Julie Marin[a–c], Michael Suleski[a–c], Madeline Paymer[a–c], and Sudhir Kumar[a–c]

[a]Center for Biodiversity, Temple University, Philadelphia, PA 19122, USA.
[b]Institute for Genomics and Evolutionary Medicine, Temple University, Philadelphia, PA 19122, USA.
[c]Department of Biology, Temple University, Philadelphia, PA 19122, USA.

[1]To whom correspondence may be addressed. E-mail: sbh@temple.edu

**Short title:** Timetree of life

*Correspondence to:*
S. Blair Hedges, Ph.D.
Center for Biodiversity
1925 N 12th Street
Temple University
Philadelphia, PA  19122-1801
E-mail: sbh@temple.edu

**ABSTRACT**

Genomic data are rapidly resolving the tree of living species calibrated to time, the timetree of life, which will provide a framework for research in diverse fields of science. Previous analyses of taxonomically restricted timetrees have found a decline in the rate of diversification in many groups of organisms, often attributed to ecological interactions among species. Here we have synthesized a global timetree of life from 2,274 studies representing 50,632 species and examined the pattern and rate of diversification as well as the timing of speciation. We found that species diversity has been mostly expanding overall and in many smaller groups of species, and that the rate of diversification in eukaryotes has been mostly constant. We also identified, and avoided, potential biases that may have influenced previous analyses of diversification including low levels of taxon sampling, small clade size, and the inclusion of stem branches in clade analyses. We found consistency in time-to-speciation among plants and animals— approximately two million years—as measured by intervals of crown and stem species times. Together, this clock-like change at different levels suggests that speciation and diversification are processes dominated by random events and that adaptive change is largely a separate process.

**INTRODUCTION**

The evolutionary timetree of life (TTOL) is needed for understanding and exploring the origin and diversity of life (Hedges and Kumar 2009; Nei 2013). For this reason, scientists have been leveraging the genomics revolution and major statistical advances in molecular dating techniques to generate divergence times between populations and species. Collectively, tens of thousands of species have been timed, and new divergence time estimates are appearing in hundreds of publications each year (*Supplementary Material Figure 1*). A global synthesis of these results will allow direct comparison of the TTOL with the fossil record and Earth history and provide new opportunities for discovery of patterns and processes that operated in the past. The TTOL is also essential for studying the multidimensional nature of biodiversity and predicting how anthropogenic changes in our environment will impact the distribution and composition of biodiversity in the future (Hoffmann et al. 2010). A robust timetree of life will provide a framework for research in diverse fields of science and medicine and a stimulus for science education. Data now exist for building a synthetic species-level TTOL of substantial size from the growing knowledge (*Supplementary Material Figure 1*).

There are challenges in synthesizing a global TTOL. The most common approach for constructing a large timetree, using a sequence alignment or superalignment is possible (Smith and O'Meara 2012; Tamura et al. 2012) but not generally practical because of data matrix sparseness. For example, genes appropriate for closely related species are unalignable at higher levels, and those appropriate for higher levels are too conserved for resolving relationships of species. Disproportionate attention to some species, such as model organisms and groups of general interest (e.g., mammals and birds), also results in an uneven distribution of knowledge. In addition, computational limits are reached for Bayesian timing methods involving more than a few hundred species (Battistuzzi et al. 2011; Jetz et al. 2012).

Here we have taken an approach to build a global TTOL by means of a data-driven synthesis of published timetrees into a large hierarchy. We have synthesized timetrees and related information in 2,274 molecular studies, which we collected and curated in a knowledgebase (Hedges et al. 2006) (*Supplementary Material*). We mapped timetrees and divergence data from those studies on a robust and conservative guidetree based on community consensus (NCBI 2013) and use those times to resolve polytomies and derive nodal times in the TTOL (*Supplementary Material Figure 2*). We present this synthesis here, for use by the community, and explore how it bears on evolutionary hypotheses and mechanisms of speciation and diversification.

**RESULTS**

**A Global Timetree of Species.** Our TTOL contains 50,632 species (Fig. 1). Nearly all (~99.5%) of the 1.9 million described species of living organisms are eukaryotes (Costello et al. 2013), and the proportion is similar (99.7%) in our TTOL (Fig. 1). The naturally unbalanced shape of the TTOL, e.g., with more eukaryote than prokaryote species, allowed us to present it in a unique spiral format to accommodate its large size. A 'timeline' can be envisioned for each species in the TTOL, based on its sequence of branching events back in time to the origin of life. For example, a timeline from humans (Fig. 2) captures evolutionary events that have received the most queries (74%) by users of the TimeTree knowledgebase (Hedges et al. 2006) (*Supplementary Material*).

Linnaean taxonomic ranks exhibit temporal inconsistencies (*Supplementary Material Figure 3*) as has been suggested in smaller surveys (Hedges and Kumar 2009; Avise and Liu 2011). For example, a class (higher rank) of angiosperm averages younger than an order (lower rank) of fungus, and an order of animal averages younger than a genus (low rank) of basidiomycete fungus. Ranks for prokaryotes are all older than corresponding ranks of eukaryotes, with a genus of eubacteria averaging $715.7 \pm 139.4$ million years ago (Ma) compared with $12.6 \pm 1.2$ Ma for a genus of eukaryotes (*Supplementary Material Figure 3*; *Supplementary Material*). These inconsistencies will cause difficulties in any scientific study comparing different types of organisms, where Linnaean rank is assumed to have a temporal equivalence.

**Diversification.** The large TTOL afforded us the opportunity to examine patterns of lineage splitting across the diversity of eukaryotes (we omit prokaryotes in our TTOL analyses because they have an arbitrary species definition). Under models of "expansion," diversity will continue to expand, either at an increasing diversification rate (hyper-expansion), the same rate (constant expansion), or decreasing rate (hypo-expansion). Saturation, on the other hand, refers to a drop in rate to zero as diversity reaches a plateau (equilibrium), possibly because of density-dependent biotic factors such as species interactions (Morlon 2014). Most recent analyses, but not all (Venditti et al. 2010; Jetz et al. 2012), have suggested that hypo-expansion is the predominant pattern in the tree of life, although there has been considerable debate as to the importance of timescales, biotic or abiotic factors, and potential biases in the analyses (Sepkoski 1984; Benton 2009; Morlon et al. 2010; Rabosky et al. 2012; Cornell 2013; Rabosky 2013).

Instead, we find that constant expansion and hyper-expansion are the dominant patterns of lineage diversification in the TTOL (Fig. 3a). Our result was statistically significant irrespective of the method used, including diversification rate tests and simulations, a coalescent approach (not conducted for the TTOL), gamma tests, a clade age-size relationship, and branch length distribution analysis (*Supplementary Material*). The rate of diversification did not decrease over the history of eukaryotes (Fig. 3; *Supplementary Material Figure 4*). The observed diversification closely matched a simulated pure birth-death model, increasing slightly during the last 200 million years (Fig. 3a–b). The terminal drop in rate at ~1 Ma is a normal characteristic of diversification plots (Etienne and Rosindell 2012) related to the taxonomic level selected for the study, in this case species. This 'taxonomic bias' (rate drop) occurs at different times (Hedges and Kumar 2009) when genera, families, or other taxa are selected for study, because lower level clades (in this case, populations destined to become species) are omitted.

The TTOL was partitioned into 58 diverse Linnaean clades for coalescence and gamma tests encompassing 44,958 total species (*Supplementary Material*). The clades were chosen to represent taxonomic diversity and a range of clade size, clade age and taxonomic sampling. Species sampling levels were high, with a median clade size of 66% of known species and 84% of the clades having 40% or more of known species. Among the 10 largest clades, only one (angiosperms) showed hypo-expansion using the coalescent method while the others were exclusively hyper-expanding. Low levels of sampling are known to bias towards hypo-expansion (Cusimano and Renner 2010; Moen and Morlon 2014), and the same clade (angiosperms) was also the most poorly sampled (5%). Also, there was no support in those clades for a decline in diversification rate using the gamma test. While some methods can account for incomplete random sampling, the non-random sampling typically used by systematists (e.g., selecting for deeply branching lineages) is not accounted for by current methods and will bias results in favor of hypo-expansion (Cusimano and Renner 2010).

Focusing on the 48 smaller clades (e.g., genera and families) of tetrapods, we found that these likewise favored the expanding models (84% of clades showing significant results) rather than the saturation model (*Supplementary Material*). Of those, most (87%) were either expanding or hyper-expanding, with only 13% hypo-expanding. Concerning the gamma test, 67% of these smaller clades did not show a decline in diversification rate through time, as was true for eukaryotes as a whole (Fig. 3; gamma statistic = 136.4, $P = 1$). Hyper-expanding clades were significantly larger (Fig. 4a; $\bar{x}$, 831 versus 91 species; $P < 0.05$) and older (Fig. 4b; $\bar{x}$, 104 versus 57 Ma; $P < 0.001$) than other clades (*Supplementary Material*). We also found that TTOL branch-length distribution fit an exponential distribution significantly better than other models (*Supplementary Material*), agreeing with an earlier study (Venditti et al. 2010) using a small data set and approach.

Diversification by expansion predicts a significant correlation between clade age and clade size (*Supplementary Material*). Therefore, we evaluate the expansion hypothesis in two densely-sampled groups: mammals (5,363 species) and birds (9,879 species). We also tested the effect of stem versus crown age (Fig. 4c) on clade sizes (*Supplementary Material*); stem age includes the time elapsed on the branch leading to the crown. We examined 1,990 clades in two separate analyses: families (153 and 113 non-nested clades for birds and mammals respectively) and genera (1,115 and 609 non-nested clades for birds and mammals respectively). In each case, the relationship was highly significant for crown age ($r = 0.43$–$0.47$, $P < 0.001$) but non-significant or weakly significant ($r = 0.02$–$0.09$, $P = 0.01$–$0.82$) using stem age (*Supplementary Material*). These results demonstrate a significant relationship between clade age and clade size in major groups of vertebrates, and provide an explanation why this was not observed in past studies, as they used stem ages (crown age + stem branch time).

Our results show that that it is best to avoid stem branch time, because the length of any (stem) branch in the tree should not be related to the time depth of the descendant node (crown age; Fig. 4d–e). Therefore, the use of stem branch time will introduce large statistical noise and make the test extremely conservative. For example, when considering every node in a timetree of species, the coefficient of variation of stem branch length relative to crown age is over 200% in the best-sampled groups, mammals (208%) and birds (224%). That noise is further weighted by the pull of the present (Nee et al. 1994), which, we determined, adds 40% time (median) to crown age at any given node (if stem age is used instead of crown age), in separate analyses of birds, mammals, and all eukaryotes. This is because the pull of the present creates longer internal branches deeper in a tree, as more lineages are pruned by extinction. Therefore, the use of stem branches in diversification analyses adds noise (variance) and gives increased weight to that noise. We believe that the stronger signal of constant expansion in our results, compared with earlier studies that have supported hypo-expansion and saturation, is in part because we have identified and avoided some biases (e.g., sampling effort, clade size, and stem age) that can impact diversification analyses.

**Speciation.** The rate of formation of new species is the primary input into the diversification of life over time. This process starts as two or more lineages split from an existing species at a node in the tree. To learn more about the timing of speciation, we assembled a separate data set of studies containing timetrees that included populations and species of eukaryotes (*Supplementary Material*). Considering a standard model of speciation based on geographic isolation, in the absence of gene flow (Sousa and Hey 2013), we estimated the time required for speciation to occur. For a given species, the time-to-speciation (TTS), after isolation, must fall between the crown and the stem age of the species (Fig. 5a). All lineages younger than the crown age of a species lead to populations that are still capable of interbreeding and, therefore, must be younger than the TTS. On the other hand, the stem age is the time when a species joins its closest relative and, therefore, it must be older than the TTS. A sampling omission of either a population or species could lower the crown age estimate or raise the stem age estimate, but the resulting interval would still contain the TTS. Although coalescence of alleles may lead to overestimates of time (Nei and Kumar 2000), those times were interpreted as population and species splits, not allelic splits, in the published studies reporting them. Also, calibrations can ameliorate the effect of allelic coalescence because they are usually based on population and species divergences, not coalescence events. We conducted a simulation to test this approach (*Supplementary Material*) finding that the true TTS is estimated precisely with a mode because of skewness of the statistical distribution and it is robust to under-sampling of lineages, either younger (populations) or older (species). We also found that the true TTS was not affected by the addition of 10% noise (intervals that do not include the TTS) and only weakly affected (2.5%) by adding a large amount (50%) of noise.

Analyses of three disparate taxonomic groups (vertebrates, arthropods, and plants) with greatly varying generation times and life histories, produced similar TTSs of approximately 2 million years (Fig. 5b). For comparison, divergence times among closely related species of prokaryotes, which are defined in practical terms (Cohan 2002), are about 50–100 times older (*Supplementary Material*). The observed similarity of TTS in diverse groups of multicellular eukaryotes suggests a model of speciation that places importance on the time that two populations are isolated (Fig. 6). If speciation is an outcome of the buildup of genic incompatibilities (GIs) between isolated populations (Coyne and Orr 2004; Matute et al. 2010), then the continuous fixation of selectively neutral mutations in two populations (Zuckerkandl and Pauling 1965; Kimura 1968) will accumulate with time and eventually lead to a number, or fraction, of GIs (here, 'S-value') that will cause postzygotic reproductive isolation (Fig. 6c). In essence, speciation—in the strictest sense—can be defined as this specific moment in time, a 'point of no return', because reproductive incompatibility and isolation is inevitable at that point. The diverging lineages will remain independent forever or until they become extinct. If they do come back into contact (Fig. 6a), they might hybridize briefly but would then undergo reinforcement, leading to prezygotic reproductive isolation. We focus here on the major model, geographic isolation, but time constraints should be similar with ecological speciation, and other models exist (Coyne and Orr 2004).

A speciation clock has been suggested previously (Coyne and Orr 1989), where GIs buildup linearly over time in postzygotic reproductive isolation. However, current evidence indicates a faster-than-linear ('snowball') rate of buildup of GIs in postzygotic reproductive isolation (Matute et al. 2010). The snowball pattern of increase in GIs is not a problem for a speciation clock as long as the time of attainment of the S-value is similar in different species. Our evidence from the TTS analysis (Fig. 6b) suggests this to be the case.

Relatively few populations will remain isolated until the point of no return because environments often change rapidly in the short term, raising and lowering barriers to gene flow (Fig. 6a). This is amplified by a preponderance of low incline landscapes (Strahler 1952), resulting in a greater impact of climate and sea level change. Therefore the buildup of GIs will be reset many times before a successful speciation event occurs (Fig. 6c). This resetting, which results in failed isolates, may explain 'barcode gaps' (Puillandre et al. 2011), which we interpret as pre-speciation loss of potential lineages. We estimate that branches between nodes in well-sampled (> 20% coverage) groups of the TTOL are, on average, 4–5 million years long using a slope method (4.92 ± 0.32 million years) and a coalescent method (4.21 ± 0.76 million years) (*Supplementary Material*). This indicates that branch time is longer than TTS, and includes a significant 'lag time,' referring to any time along a branch that is not part of a successful speciation event, such as the resetting of the speciation clock by reversal of the buildup of GIs. The high variance of branch length in trees (Fig. 4d–e) may reflect the stochastic nature of environmental change and isolation of populations, contributing to a 'diversification clock' (Fig. 3). This is consistent with the absence of a strong correlation between reproductive isolation and diversification rate in some taxa (Rabosky and Matute 2013).

In relating this speciation model to a phylogeny (Fig. 6d), it is evident that the split of two lineages in a tree is not the speciation event but rather the moment of isolation of two or more populations that will remain isolated until the point of no return. For example, if we assume a TTS of 2 million years, then two populations that only recently became species would have split ~2 Ma (Fig. 6d, species 1–2). Yet, much deeper in a tree, where branches may be tens of millions of years long, the TTS is a relatively small fraction of total branch time, which makes a splitting event more-or-less equivalent to the speciation event. Furthermore, branch splitting leading to eventual species formation can happen simultaneously as long as the resulting populations remain isolated until the point of no return (Fig. 6d, species 5–9).

**DISCUSSION**

These results have implications for patterns of species diversification and the interpretation of timetrees. If adaptation is largely decoupled from speciation, we should not expect to see major diversification rate increases following mass extinction events, even though large adaptive changes took place at those times. That expectation is realized in our analyses (Fig. 3) where we see constant splitting through time across the two major Phanerozoic extinction events (251 and 66 Ma). Likewise, our diversification analyses of smaller groups, such as birds and mammals (*Supplementary Material Figure 5*), and past studies of those groups (Bininda-Emonds et al. 2007; Meredith et al. 2011; Jetz et al. 2012), have not found rate increases immediately following the end-Cretaceous mass extinction event (66 Ma). Rate decreases from the extinction events themselves are not expected because of the inability to detect, in trees, a large proportion of species extinctions that occurred (Nee and May 1997).

These results, which show a consistency in TTS among groups, infers that the time-based acquisition of genic incompatibilities is driving reproductive isolation, and not adaptive change. By implication, geographically isolated populations, even if morphologically different and diagnosable, are expected to be species until they have reached the point of no return. This is because some differentiation and adaptive change should occur in isolation, and many isolates will be ephemeral, merging with other isolates or disappearing and never becoming species that enter the tree of life. More population data are needed before it is possible to identify the point of no return with precision. Nonetheless, these data suggest that, in most cases, described species separated by only tens of thousands of years are not real species. The Linnaean rank of subspecies, which has declined in use for decades, might be appropriate for such diagnosable isolates that have not yet reached the point of no return. Over-splitting of species by taxonomists may also explain the large rate increase (hyper-expansion) in the last ten million years of eukaryote history (Fig. 3b), although other explanations (e.g., climate change) cannot be ruled out.

In summary, the diversification of life as a whole is expanding at a constant rate and only in some small clades (< 500 species) is there evidence of decline or saturation in diversification rate. We believe these results can be explained as many different groups of organisms undergoing expansion and contraction through time, with those patterns captured in different stages (expanding, slowing, or in equilibrium). If true, the predominant pattern of expansion in large clades is expected from the law of large numbers, where such smaller, random events, would average to a constant rate. Constant expansion also follows from random environmental changes leading to isolation and speciation. Rate constancy is consistent with the fossil record (Benton 2009) and does not deny the importance of biotic factors in evolution (Ricklefs 2007), but it suggests an uncoupling of speciation and adaptation. Cases where the phenotype has changed little (e.g., cryptic species) or greatly during the time-to-speciation are interpreted here as evidence of uncoupling. The lineage splitting seen in trees probably reflects, in most instances, random environmental events leading to isolation of populations, and potentially many in a short time. However, the relatively long time-to-speciation (~2 million years), a process resulting from random genetic events, will limit the number of isolates that eventually become species. Under this model, diversification is the product of those two random processes, abiotic and genetic, and rate increases (bursts of speciation) are more likely to be associated with long-term changes in the physical environment (e.g., climate, sea level) causing extended isolation rather than with short-term ecological interactions. Adaptive change that characterizes the phenotypic diversity of life would appear to be a separate process from speciation and diversification. Although a full understanding of these processes remains a challenge, determining how speciation and adaptation are temporally related—coupled or uncoupled—would be an important 'next step.'

**Materials and Methods**

**Detailed methods are described in *Supplementary Material*.**

**Timetree of life (TTOL) data collection.** We synthesized the corpus of scientific literature where the primary research on the timetree of life is published. We first identified and collected all peer-reviewed publications in molecular evolution and phylogenetics that reported estimates of time of divergence among species. These included phylogenetic trees scaled to time (timetrees) and occasionally tables of time estimates and regular text. We assembled timetree data from 2,274 studies (http://www.timetree.org/reference_list.php) that have been published between 1987 and April, 2013, as well as two timetrees estimated herein (*Supplementary Material Table 1*). Most (96%) of nodal times used were published in the last decade.

**Timetree of life (TTOL) analytics and synthesis.** The fundamental unit of synthesis in our database was the divergence time of a pair of clades (A and B) which have directly descended from an ancestor (X) in the tree of life. Timetree data enables synthesis across studies without taking a traditional phylogeny-partition-based supertree approach, which is not feasible because of the extreme sparseness of the species-studies data. We used a Hierarchical Average Linkage (HAL) method of estimating divergence times (*T*'s) of clade pairs to build a Super Timetree (STT), along with a procedure for testing and updating topological partitions to ensure the highest degree of consistency with individual timetrees in every study. For the TTOL, uncertainty derived from individual studies is available for each node (*Supplementary Material Table 2*). Branch time modes of different Linnaean categories were estimated (*Supplementary Material Table 3*).

**Diversification analyses.** All diversification analyses were performed in R (http://www.r-project.org/) with the APE package (Paradis et al. 2013). The gamma statistic was employed to detect decrease in speciation rate over the history of the tree. In order to test if diversification is saturated or expanding, we used a coalescent method as well as the relationship between the ages of different clades and their sizes (numbers of species). The number and location of shifts in diversification rates were also tested. We also estimated among-study uncertainty in the both the LTT curve and rate test by producing 500 replicates of the TTOL, sampling actual study times at each node randomly. These replicates were then used in the analysis of rate shifts to estimate and test rate change. Results of all diversification analyses were summarized (*Supplementary Material Tables 4–10*).

**Time-to-speciation analyses.** In addition to our species-level TTOL data collection described above, we collected a separate data set on TTS from published molecular timetrees that included timed nodes among populations and closely related species of three major groups: vertebrates, arthropods, and plants (*Supplementary Material Tables 11–13*). To test the robustness of our approach for estimating TTS, we used simulations. A birth-death tree was simulated using the function 'sim.bd.taxa' (TreeSim in R).

References

Avise, JC, JX Liu. 2011. On the temporal inconsistencies of Linnean taxonomic ranks. Biological Journal of the Linnean Society 102:707-714.

Battistuzzi, FU, P Billing-Ross, A Paliwal, S Kumar. 2011. Fast and slow implementations of relaxed-clock methods show similar patterns of accuracy in estimating divergence times. Molecular Biology and Evolution 28:2439-2442.

Benton, MJ. 2009. The Red Queen and the Court Jester: species diversity and the role of biotic and abiotic factors through time. Science 323:728-732.

Bininda-Emonds, ORP, M Cardillo, KE Jones, RDE MacPhee, RMD Beck, R Grenyer, SA Price, RA Vos, JL Gittleman, A Purvis. 2007. The delayed rise of present-day mammals. Nature 446:507-512.

Cohan, FM. 2002. What are bacterial species? Annual Review of Microbiology 56:457-487.

Cornell, HV. 2013. Is regional species diversity bounded or unbounded? Biological Reviews 88:140-165.

Costello, MJ, RM May, NE Stork. 2013. Can we name Earth's species before they go extinct? Science 339:413-416.

Coyne, JA, HA Orr. 1989. Patterns of Speciation in *Drosophila*. Evolution 43:362-381.

Coyne, JA, HA Orr. 2004. Speciation. Sunderland, Massachusetts: Sinauer Associates, Inc.

Cusimano, N, SS Renner. 2010. Slowdowns in Diversification Rates from Real Phylogenies May Not be Real. Systematic Biology 59:458-464.

Etienne, RS, J Rosindell. 2012. Prolonging the past counteracts the pull of the present: protracted speciation can explain observed slowdowns in diversification. Syst Biol 61:204-213.

Hedges, SB, J Dudley, S Kumar. 2006. TimeTree: a public knowledge-base of divergence times among organisms. Bioinformatics 22:2971-2972.

Hedges, SB, S Kumar, editors. 2009. The timetree of life. Oxford: Oxford University Press.

Hoffmann, MC Hilton-TaylorA Angulo, et al. 2010. The impact of conservation on the status of the world's vertebrates. Science 330:1503-1509.

Jetz, W, GH Thomas, JB Joy, K Hartmann, AO Mooers. 2012. The global diversity of birds in space and time. Nature 491:444-448.

Kimura, M. 1968. Evolutionary rate at the molecular level. Nature 217:624-626.

Matute, DR, IA Butler, DA Turissini, JA Coyne. 2010. A Test of the Snowball Theory for the Rate of Evolution of Hybrid Incompatibilities. Science 329:1518-1521.

Meredith, RW, JE Janecka, J Gatesy, et al. 2011. Impacts of the Cretaceous Terrestrial Revolution and KPg Extinction on Mammal Diversification. Science 334:521-524.

Moen, D, H Morlon. 2014. Why does diversification slow down? Trends Ecol Evol 29:190-197.

Morlon, H. 2014. Phylogenetic approaches for studying diversification. Ecology Letters 17:508-525.

Morlon, H, MD Potts, JB Plotkin. 2010. Inferring the dynamics of diversification: a coalescent approach. Plos Biology 8.

NCBI. 2013. Taxonomy Browser. http://www.ncbi.nlm.nih.gov/. Bethesda, Maryland, U.S.A.: National Center for Biotechnology Information.

Nee, S, RM May. 1997. Extinction and the loss of evolutionary history. Science 278:692-694.

Nee, S, RM May, PH Harvey. 1994. The reconstructed evolutionary process. Philosophical Transactions of the Royal Society of London Series B-Biological Sciences 344:305-311.

Nei, M. 2013. Mutation driven evolution. New York: Oxford University Press.

Nei, M, S Kumar. 2000. Molecular Evolution and Phylogenetics. New York: Oxford University Press.

Paradis, E, B Bolker, J Claude, et al. 2013. Ape: Analyses of phylogenetics and evolution. http://CRAN.R-project.org/package=ape. Vienna, Austria: Comprehensive R Archive Network.

Puillandre, N, A Lambert, S Brouillet, G Achaz. 2011. ABGD, Automatic Barcode Gap Discovery for primary species delimitation. Molecular Ecology 21:1864-1877.

Rabosky, DL. 2013. Diversity-dependence, ecological speciation, and the role of competition in macroevolution. Annual Review of Ecology, Evolution, and Systematics 44:481-502.

Rabosky, DL, DR Matute. 2013. Macroevolutionary speciation rates are decoupled from the evolution of intrinsic reproductive isolation in Drosophila and birds. Proceedings of the National Academy of Sciences of the United States of America 110:15354-15359.

Rabosky, DL, GJ Slater, ME Alfaro. 2012. Clade age and species richness are decoupled across the Eukaryotic tree of life. Plos Biology 10:e1001381.

Ricklefs, RE. 2007. Estimating diversification rates from phylogenetic information. Trends in Ecology & Evolution 22:601-610.

Sepkoski, JJ. 1984. A kinetic model of Phanerozoic taxonomic diversity. 3. Post-Paleozoic families and mass extinctions. Paleobiology 10:246-267.

Smith, SA, BC O'Meara. 2012. treePL: divergence time estimation using penalized likelihood for large phylogenies. Bioinformatics 28:2689-2690.

Sousa, V, J Hey. 2013. Understanding the origin of species with genome-scale data: modelling gene flow. Nature Reviews Genetics 14:404-414.

Strahler, AN. 1952. Hypsometric (area-altitude) analysis of erosional topography. Geological Society of America Bulletin 63:1117-1142.

Tamura, K, FU Battistuzzi, P Billing-Ross, O Murillo, A Filipski, S Kumar. 2012. Estimating divergence times in large molecular phylogenies. Proceedings of the National Academy of Sciences of the United States of America 109:19333-19338.

Venditti, C, A Meade, M Pagel. 2010. Phylogenies reveal new interpretation of speciation and the Red Queen. Nature 463:349-352.

Zuckerkandl, E, L Pauling. 1965. Evolutionary divergence and convergence in proteins. In: V Bryson, HJ Vogel, editors. Evolving Genes and Proteins. New York: Academic Press. p. 97-165.

**ACKNOWLEDGEMENTS**

The work was supported by grants from the U.S. National Science Foundation (0649107, 0850013, 1136590, 1262481, 1262440), the Science Foundation of Arizona, the U.S. National Institutes of Health (HG002096-12), and from the NASA Astrobiology Institute (NNA09DA76A). We thank K. Boccia, J.R. Dave, S.L. Hanson, A. Hippenstiel, M. McCutchan, Y. Plavnik, A. Shoffner, and L.-W. Wu for database assistance; K.R. Hargreaves, S.G. Loelius, and K.M. Wilt for population data collection assistance; B.A. Gattens, A.Z. Kapinus, L.M. Lee, A.B. Marion, J. McKay, A. Nikolaeva, M.J. Oberholtzer, M.A. Owens, V.L. Richter, and L. Stork for illustration assistance; R. Chikhi for programming assistance; J. Hey for comments on the manuscript; Penn State and Arizona State universities; and authors of studies for contributing timetrees.

**Figure legends**

**Fig. 1. (2 columns)** A timetree of 50,632 species synthesized from times of divergence published in 2,274 studies. Evolutionary history is compressed into a narrow strip and then arranged in a spiral with one end in the middle and the other on the outside. Therefore, time progresses across the width of the strip at all places, rather than along the spiral. Time is shown in billions of years on a log scale and indicated throughout by bands of gray. Major taxonomic groups are labeled.

**Fig. 2**. (**1 column**) A timeline from the perspective of humans, showing divergences with other groups of organisms. In each case, mean ± standard error (among studies) is shown, with number of studies in parentheses. Also shown are the times for the origin of life, eukaryotes, and last universal common ancestor (Hedges and Kumar 2009).

**Fig. 3. (1 column)** Patterns of lineage diversification. (*a*) Cumulative lineages-through-time (LTT) curve for eukaryotes (50,455 sp.), in black, showing the number of lineages through time (unsmoothed, dashed; smoothed, solid) and variance (red, 500 replicates). (*b*) same LTT curve (black line), but compared with a simulated constant-expansion LTT curve ($\lambda = 0.073$ and $\mu = 0.070$) shown as ± 99% confidence intervals (red). (*c*) Diversification rate plot of same data showing only significant changes in rate as determined in maximum-likelihood tests; variance (red, 500 replicates) shown as ± 99% intervals.

**Fig. 4. (1 column)** Diversification analyses of major Linnaean clades in the timetree of life. (*a–b*) Results of coalescent analyses testing models of diversification. Of 48 tetrapod clades, 37 showed significant model selected and they were used in these analyses. (*a*) Effect of clade size (number of described species). (*b*) Effect of clade age. (*c*) Diagram illustrating difference between stem and crown age for two clades. (*d–e*) Relationship of stem branch and crown age in mammals (*d*, $r^2 = 0.07$) and birds (*e*, $r^2 = 0.04$).

**Fig. 5. (2 columns)** Estimation of time-to-speciation. (*a*) Analytical design showing expected results. Intervals between stem age and crown age of seven species contain time-to-speciation (dashed line). Divergences among populations, and corresponding histogram, are in red. (*b*) Colored histograms of observed time-to-speciation, showing modes (vertical lines) and confidence intervals (bars) in vertebrates (N = 213, 2.1 Ma, 1.74–2.55 Ma), arthropods (N = 85, 2.2 Ma, 1.57–3.07 Ma), and plants (N = 55, 2.7 Ma, 2.37–3.63 Ma). Black curves to right are histograms of population divergences.

**Fig. 6. (1 column)** Summary model of speciation. (*a*) Biogeographic history, showing the contact and isolation of areas occupied by the two populations. (*b*) Phylogenetic lineages, showing times of independence (two lineages) and times of interbreeding (one lineage). (*c*) Genic incompatibilities, GIs, between the two populations, showing how they accrue at a time-dependent rate during geographic isolation, reset to zero during contact (interbreeding), increase to the S-value (the number of GIs that will cause speciation, the point of no return), and continue increasing beyond the S-value despite later contact of the newly-formed species. (*d*) Hypothetical phylogeny, with numbered species, illustrating parameters of speciation in (*a–c*) to splits and branches in a tree.

Fig 1

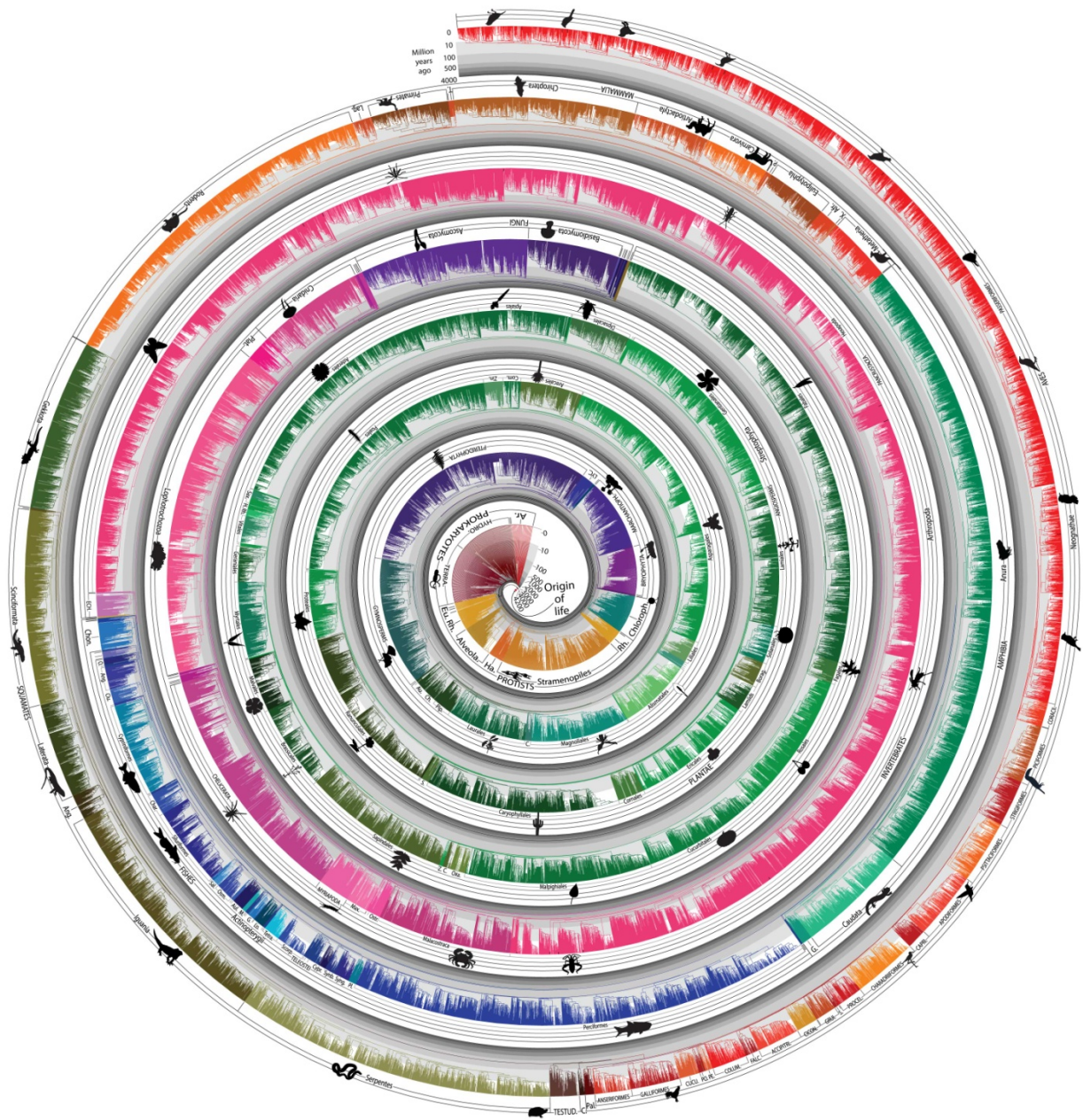

Fig 2

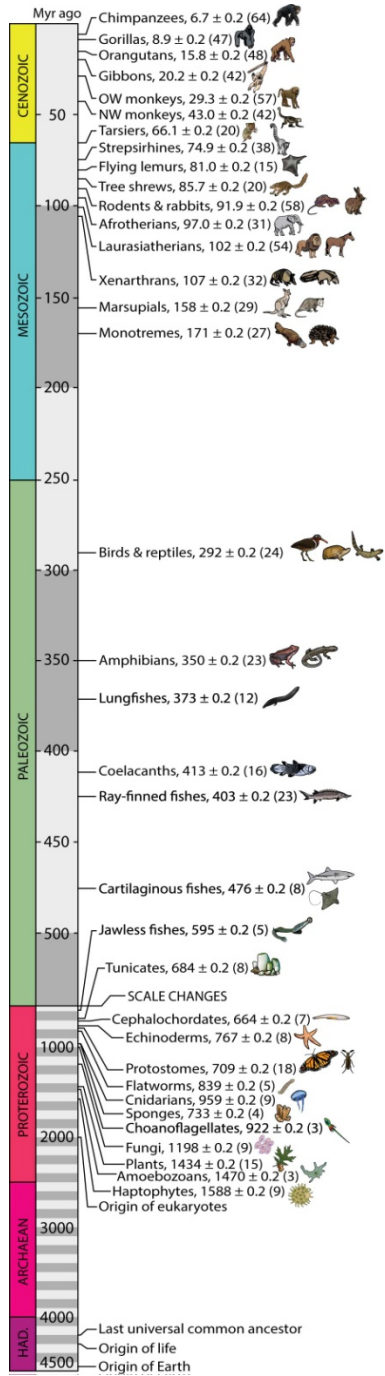

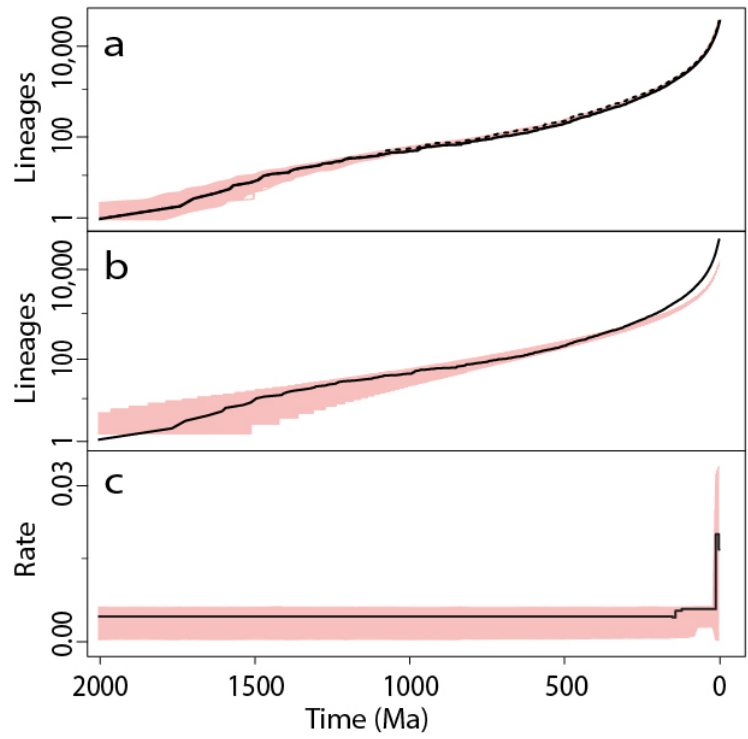

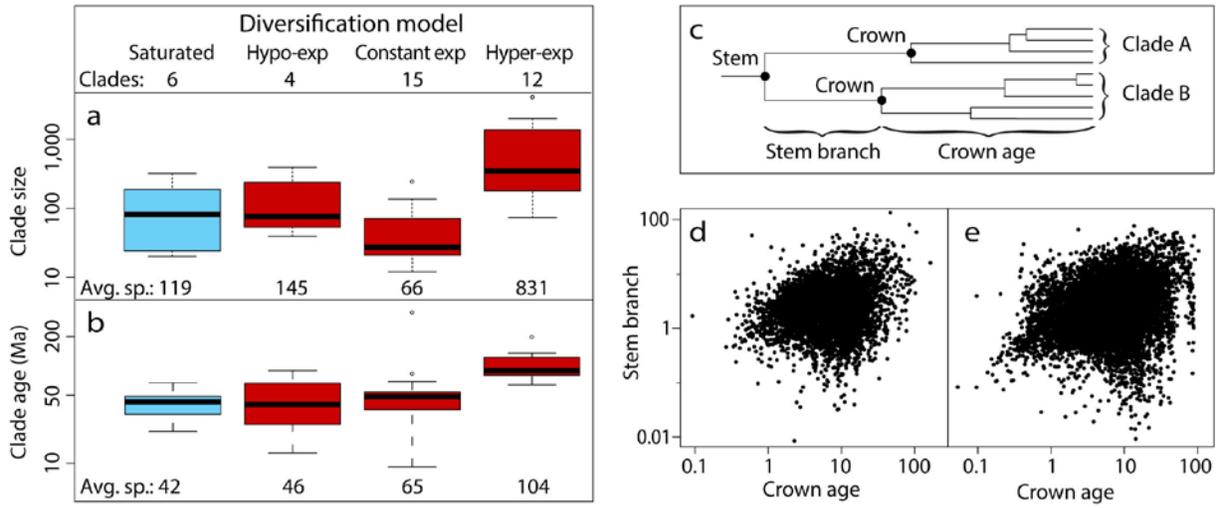

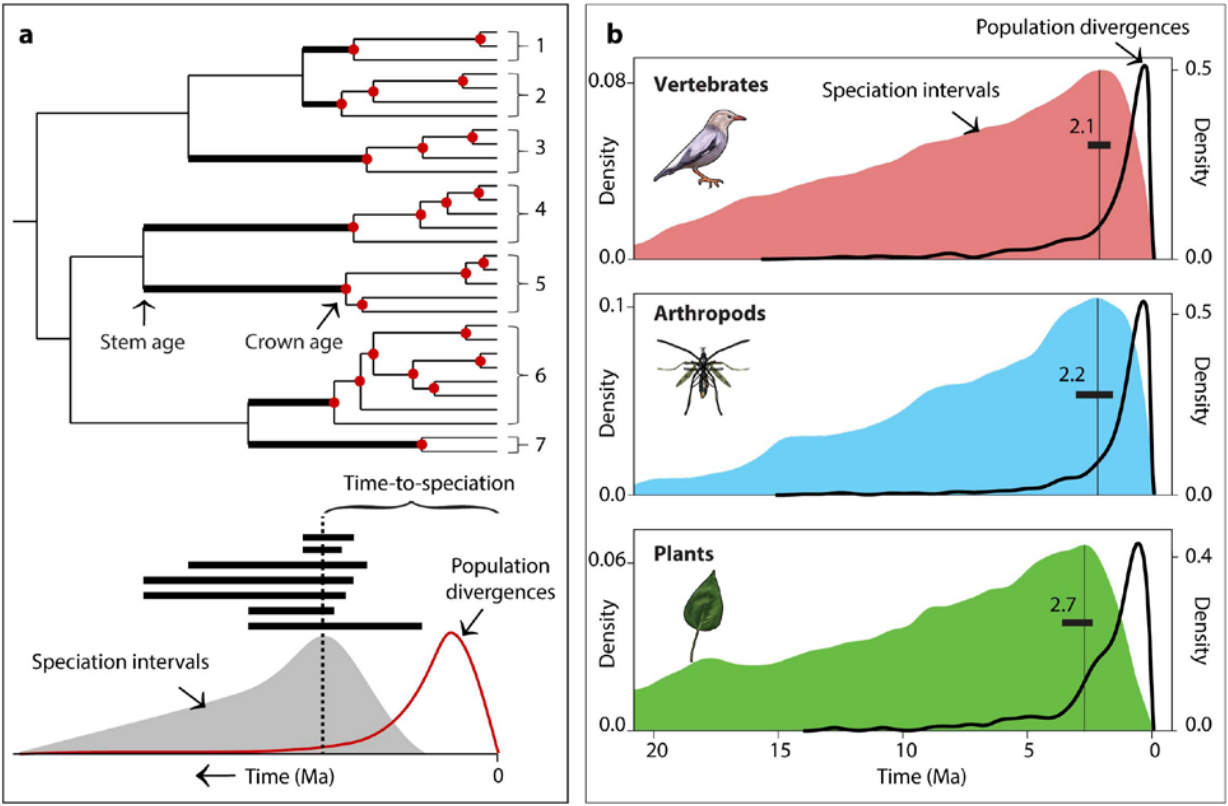

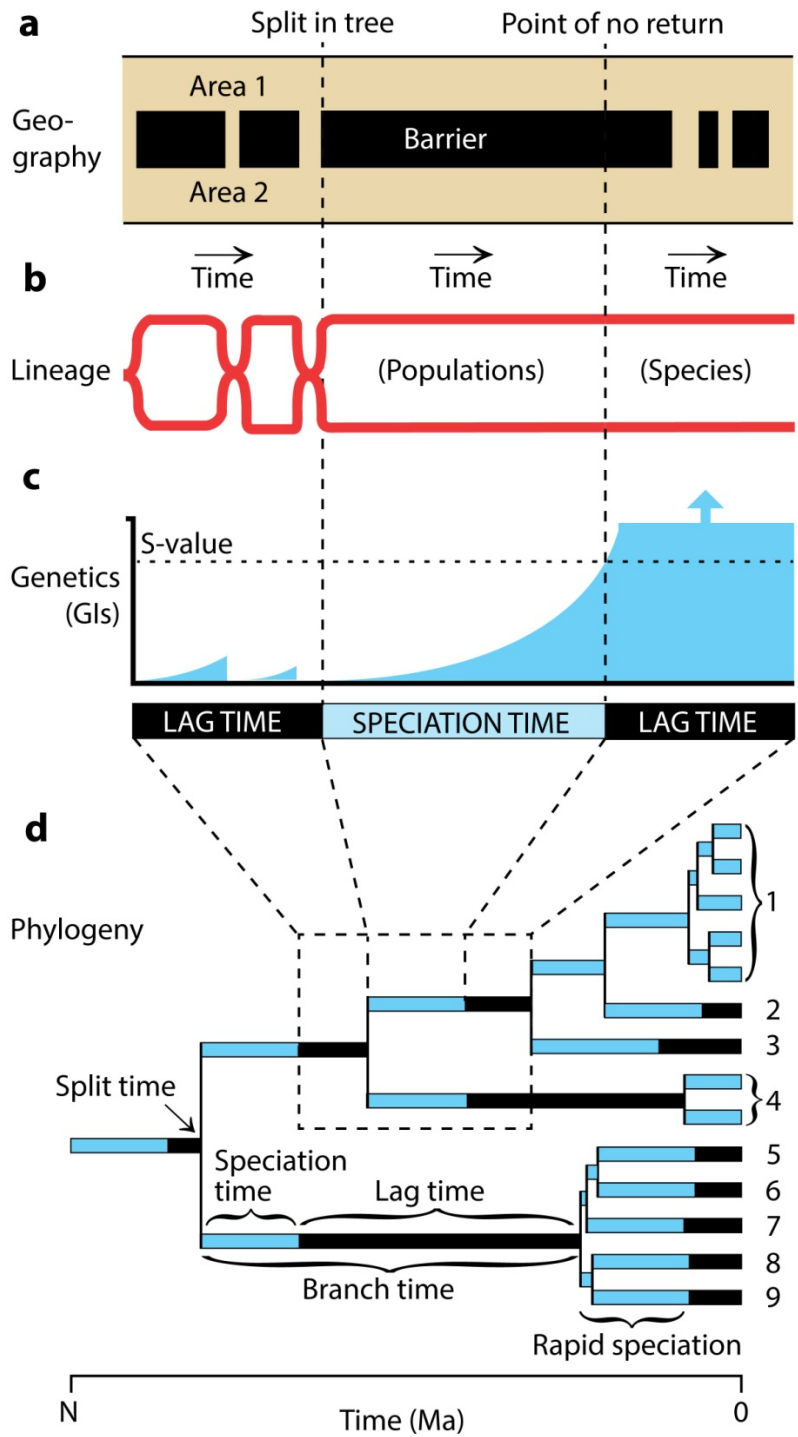